\documentclass[rnote]{aa}

\usepackage{txfonts}
\usepackage{graphicx}

\usepackage{natbib}

\def\rxte{{\it RXTE}}

\def\lan{ \langle}
\def\ran{ \rangle}

\def\mdot{\ifmmode \dot M \else $\dot M$\fi}    % accretion rate
\def\mxd{\ifmmode \dot {M}_{x} \else $\dot {M}_{x}$\fi}
\def\med{\ifmmode \dot {M}_{Edd} \else $\dot {M}_{Edd}$\fi}
\def\bff{\ifmmode B_{f} \else $B_{f}$\fi}

\def\apj{\ifmmode ApJ \else ApJ \fi}    % lower
\def\apjl{\ifmmode  ApJ \else ApJ \fi}    %
\def\aap{\ifmmode A\&A \else A\&A\fi}    %
\def\mnras{\ifmmode MNRAS \else MNRAS \fi}    %
\def\nat{\ifmmode Nature \else Nature \fi}

\def\prl{\ifmmode Phys. Rev. Lett. \else Phys. Rev. Lett.\fi}
\def\prd{\ifmmode Phys. Rev. D. \else Phys. Rev. D.\fi}

\def\ms{\ifmmode {\rm M_{\odot}} \else ${\rm M_{\odot}}$\fi}    % lower
\def\na{\ifmmode \nu_{\rm A} \else $\nu_{\rm A}$\fi}    % Alfven frequency
\def\nk{\ifmmode \nu_{\rm k} \else $\nu_{\rm k}$\fi}    % Keplerian frequency
\def\ns{\ifmmode \nu_{{\rm s}} \else $\nu_{{\rm s}}$\fi}
\def\no{\ifmmode \nu_{1} \else $\nu_{1}$\fi}    % lower
\def\nt{\ifmmode \nu_{2} \else $\nu_{2}$\fi}    % upper
\def\ntk{\ifmmode \nu_{\rm 2k} \else $\nu_{\rm 2k}$\fi}    % upper
\def\dnmax{\ifmmode \Delta \nu_{\rm max} \else $\Delta \nu_{\rm max}$\fi}
\def\dnobs{\ifmmode \Delta \nu_{\rm obs} \else $\Delta \nu_{\rm obs}$\fi}
\def\ntmax{\ifmmode \nu_{\rm 2max} \else $\nu_{\rm 2max}$\fi}    %
\def\xmax{\ifmmode {\rm X_{\rm max}} \else ${\rm X_{max}}$\fi}

\def\ntobs{\ifmmode \nu_{\rm 2obs} \else $\nu_{\rm 2obs}$\fi}    %
\def\nomax{\ifmmode \nu_{\rm 1max} \else $\nu_{\rm 1max}$\fi}    % upper
\def\nn{\ifmmode \nu_{\rm NBO} \else $\nu_{\rm NBO}$\fi}    % HBO
\def\nh{\ifmmode \nu_{\rm HBO} \else $\nu_{\rm HBO}$\fi}    % HBO
\def\nqpo{\ifmmode \nu_{QPO} \else $\nu_{QPO}$\fi}    % HBO
\def\nz{\ifmmode \nu_{o} \else $\nu_{o}$\fi}    % HBO
\def\nht{\ifmmode \nu_{H2} \else $\nu_{H2}$\fi}    % HBO
\def\ns{\ifmmode \nu_{s} \else $\nu_{s}$\fi}    %
\def\nb{\ifmmode \nu_{{\rm b}} \else $\nu_{{\rm b}}$\fi}
\def\nkm{\ifmmode \nu_{km} \else $\nu_{km}$\fi}    %
\def\ka{\ifmmode \kappa \else \kappa\fi}    %
\def\dn{\ifmmode \Delta\nu \else \Delta\nu\fi}

\def\vk{\ifmmode v_{\rm k} \else $v_{\rm k}$\fi}    % Keplerian velocity
\def\va{\ifmmode v_{\rm A} \else $v_{\rm A}$\fi}    % Alfven velocity
\def\vf{\ifmmode v_{\rm ff} \else $v_{\rm ff}$\fi}    % free fall velocity

\def\rs{\ifmmode R_{\rm s} \else $R_{\rm s}$\fi}    %
\def\ra{\ifmmode R_{\rm A} \else $R_{\rm A}$\fi}    % Alfven radius
\def\rso{\ifmmode R_{S1} \else $R_{S1}$\fi}    % sonic point radius
\def\rst{\ifmmode R_{S2} \else $R_{S2}$\fi}    % sonic point radius
\def\rmm{\ifmmode {\rm R_{M}} \else ${\rm R_{M}}$\fi}    %
\def\rco{\ifmmode {\rm R_{co}} \else ${\rm R_{co}}$\fi}    %
\def\ris{\ifmmode {\rm R}_{{\rm ISCO}} \else $ {\rm R}_{{\rm ISCO}} $\fi}
\def\rsix{\ifmmode R_{6} \else $R_{6}$\fi}

\def\sax{\ifmmode {SAX J1808.4-3658} \else  SAX J1808.4-3658\fi}
\def\x1807{\ifmmode {XTE J1807-294} \else XTE J1807-294\fi}
\def\1739{\ifmmode {XTE  J1739-285} \else XTE  J1739-285\fi}
 \def\exo{\ifmmode {EXO 0748-676} \else EXO 0748-676\fi}

\begin{document}

\title{The correlations between the spin frequencies and kHz QPOs of
Neutron Stars in LMXBs}

\author{H.X. Yin$^{1}$, C.M. Zhang$^{1}$, Y.H. Zhao$^{1}$, Y.J. Lei$^{2}$,
 J.L. Qu$^{2}$, L.M. Song$^{2}$, F. Zhang$^{2}$ }

\institute{1. National Astronomical Observatories,
 Chinese Academy of Sciences, Beijing 100012, China\\
 2. Astronomical Institute, Institute of High Energy Physics,
 Chinese Academy of Sciences, Beijing 100039, China\\}

\offprints{zhangcm@bao.ac.cn}
\authorrunning{H.X. Yin et al. }
\titlerunning{Correlations between the spin frequencies and kHz QPOs}
%\date{Received date ; accepted date}

\abstract
%content
{}
%aims
{We studied the correlations between spin frequencies and kilohertz
quasi-periodic oscillations (kHz QPOs) in neutron star low mass
X-ray binaries.}
%methods
{The updated data of kHz QPOs and spin frequencies are statistically
analyzed.}
%results
{We found that when two simultaneous kHz QPOs are present in the
power spectrum, the minimum frequency of upper kHz QPO is at least
1.3 times larger than the spin frequency,  i.e.
$\nu_{s}<\nu_{2min}/1.3$. We also found that the average kHz QPO
peak separation in 6 Atoll sources anti-correlates with the spin
frequency in the form
$\lan\dn\ran=-(0.19\pm0.05)\ns+(389.40\pm21.67)$Hz. If we shifted
this correlation in the direction of the peak separation by a factor
of 1.5, this correlation matches the data points of the two
accretion powered millisecond X-ray pulsars, SAX J1808.4-3658 and
XTE J1807-294. }
%conclusion
{}  \maketitle

\section{Introduction}
Since the launch of the Rossi X-Ray Timing Explorer (\rxte) ten
years ago, kilohertz quasi-periodic oscillations (kHz QPOs) have
been detected in about thirty neutron star low mass X-ray binaries
(NS/LMXBs; see van der Klis 2006, for a recent review). The QPOs
often occur in pairs, the upper-frequency ($\nt$) and the
lower-frequency ($\no$).
These kHz QPOs appear in four categories of NS/LMXBs, i.e. the
bright Z sources, the less luminous Atoll sources (see Hasinger \&
van der Klis 1989 for the definition of Atoll and Z classes),
accretion-powered millisecond X-ray pulsars (AMXPs) and other
unidentified sources (see e.g., van der Klis 2006 and references
therein). The kHz QPOs and other observed characteristic frequencies
in these sources follow tight correlations among each other (e.g.,
Psaltis et al. 1998, 1999ab; Stella et al. 1999; Belloni et al.
2002, 2005, 2007; Zhang et al. 2006a).

A 401 Hz coherent pulsation, and a near 401 Hz X-ray burst
oscillation frequency are found in SAX J1808.4-3658 (Chakrabarty et
al. 2003; Wijnands et al. 2003), suggesting that the burst frequency
is equal to the spin frequency ($\nu_{s}$) in this object
 (e.g., Strohmayer \& Bildsten 2003; Wijnands et al. 2003;
Muno 2004).
 In some sources showing both twin kHz QPOs and spin
frequencies, the peak separation ($\Delta\nu=\nt-\no$) is generally
inconsistent with being equal to the spin frequency (e.g. M\'{e}ndez
\& van der Klis 1999; Jonker, M\'{e}ndez, \& van der Klis 2002a).
But the ratio between  the peak separation and spin frequency
clusters at around $\sim$1 or $\sim$0.5 (e.g., Wijnands et al. 2003;
Wijnands 2005; Linares et al. 2005; Zhang et al. 2006b).
In this research note, we study the relation between kHz QPOs and
spin frequencies.

\section{Correlations between Spins and kHz QPOs}
From 35 NS/LMXBs with the kHz QPOs and/or  spin frequencies, 21(6)
sources show twin (single) kHz QPOs, and 22 sources show spin or/and
burst frequencies (7 spin and 17 burst sources; see Tab. 1).
%

%-----------------------------------------------------------------
\begin{table*}[t]
\begin{center}
  \caption{\bf List of LMXBs with the simultaneously detected twin kHz QPOs or  spin frequencies.}
  \label{Tab:1}
  \begin{tabular}{lccccccl}
  \hline
  \hline
   Sources          & $\nu$$_{1}$$^{(1)}$   & $\nu$$_{2}$$^{(2)}$ & $\Delta$$\nu$$^{(3)}$ & $\nu$$_{2}$/$\nu$$_{1}$$^{(4)}$ & $\nu$$_{burst}$$^{(5)}$ & $\nu$$_{pulse}$$^{(6)}$ & References   \\
       & (Hz) & (Hz) & (Hz) & & (Hz) & (Hz) & \\
  \hline
{\bf Millisecond pulsars (7)}                 \\
  \hline
IGR J00291+5934    &-    &    -      &   -        &     -      &      -         &  599     &      1                     \\
XTE J0929-314      &-    &    -      &   -        &     -      &     -          &  185     &      K                     \\
XTE J1751-305      &-    &    -      &   -        &     -      &     -          &  435     &      K                     \\
XTE J1807-294      &127-360 & 353-587 & 179-247 & 1.51-2.78  &     -          &  191     &      2,3                    \\
SAX J1808.4-3658   & 499   &    694   &  195       &  1.39  &     401        &  401     &     K,4                     \\
XTE J1814-338      &-    &    -      &   -        &     -      &     314        &  314     &      K                     \\
HETE J1900.1-2455$^{a}$  &-    &    -      &   -        &     -      &      -         &  377   &      5                     \\
  \hline
{\bf Z sources (8)}                \\
  \hline
Sco X-1     &  544-852  & 844-1086  &223-312     &  1.26-1.57    &   -    &     649$^{\dag}$         &   M,B,K                 \\
GX 340+0    &   197-565 & 535-840   &275-413     &  1.49-2.72    &   -    &     412$^{\dag}$         &   B,K,P,6                \\
GX 349+2    &   712-715 & 978-985   &266-270     &  1.37-1.38    &   -    &     752$^{\dag}$         &   B,K,7                \\
GX 5-1      &  156-634  & 478-880   &232-363     &  1.38-3.06    &   -    &     368$^{\dag}$         &   B,K,P,8                \\
GX 17+2     &  475-830  & 759-1078  &233-308     &  1.28-1.60    &   -    &     584$^{\dag}$         &   B,K,P,9                \\
Cyg X-2     &     532   &  856      & 324        &   1.61        &   -    &     658$^{\dag}$         &   B,K,P                  \\
Cir X-1     & 56-226    & 229-505   & 173-340    &  2.23-4.19    &-       &     176$^{\dag}$         &    10\\
XTE J1701-462 &  620    & 909       & 289        & 1.47          &-       &     699$^{\dag}$         &  11\\
  \hline
{\bf Atoll sources (16)}               \\
  \hline
4U 0614+09  &  153-823  & 449-1162  & 238-382    &   1.38-2.93   &    -    &    345$^{\dag}$         &   B,K,P,12,13             \\
XB 1254-690 &   -       &   -       &   -        &     -         &  95     &    -         & 14\\
4U 1608-52  &  476-876  & 802-1099  &224-327     &  1.26-1.69    &  619    &    -         &   M,B,K,15                 \\
4U 1636-53  &  644-921  & 971-1192  &217-329     &  1.24-1.51    &  581    &    -         &   B,K,P,16,17              \\
4U 1702-43  &  722      &  1055     & 333        &     1.46      &  330    &    -         &   K,P,18                  \\
4U 1705-44  &  776      &  1074     & 298        &     1.38      &   -     &    826$^{\dag}$         &   B,K,P             \\
4U 1728-34  &  308-894  & 582-1183  &271-359     &  1.31-1.89    &  363    &    -         &   B,K,P,13,19       \\
KS 1731-260 &  903      &  1169     & 266        &    1.29       &  524    &    -         &   B,K,P          \\
4U 1735-44  & 640-728   & 982-1026  &296-341     &  1.41-1.53    &   -     &    755$^{\dag}$         &   B,K,P             \\
XTE J1739-285$^{a}$ & - & -   &   -  &   -    & 1122 & - &20\\
A 1744-361$^{a}$  &   -       &   -       &   -        &     -         &    530  &  -           &   21 \\
SAX J1750.8-2900$^{a}$ &-     &   -       &   -        &      -        &  601    &    -         &   K,22                      \\
4U 1820-30  &  790      &  1064     & 273        &    1.35       &   -     &    818$^{\dag}$         &   B,K,P                  \\
Aql X-1$^{a}$          &-     &   -       &   -        &      -        &  549    &    -         &   B,K,P                     \\
4U 1915-05  & 224-707   & 514-1055  &290-353     &  1.49-2.3     &  270    &    -         &   B,K,P             \\
XTEJ2123-058& 849-871   &1110-1140  &261-270     &  1.31-1.31    &  -      &    854$^{\dag}$         &   B,K,P                  \\
  \hline
{\bf Other sources (4)}                         \\
  \hline
EXO 0748-676$^{a}$    &  -  &   -       &   -        &     -         &  45     & -            & K,23\\
MXB 1659-298    &  -    &   -       &    -       &      -        &  567    &    -         &    K,24                    \\
MXB 1743-29     &  -    &   -       &    -       &      -        &  589    &    -         &    K,25                    \\
SAX J1748.9-2021&  -    &   -       &    -       &      -        &  410    &    -         &    K,26                     \\
  \hline
  \hline
  \end{tabular}
\begin{tabular}{p{\linewidth}}
{ $^{a}$:sources with only a single QPO detected. $^{\dag}$:the
inferred upper limit of the spin frequency using the relation
$\nu_{s}<\nu_{2min}/1.3$ (see text). $^{(1)}$: lower-frequencies;
$^{(2)}$: upper-frequencies; $^{(3)}$: separations of twin kHz QPOs;
$^{(4)}$: ratios between the upper- and lower-frequencies; $^{(5)}$:
burst frequency $\nu$$_{burst}$; $^{(6)}$: coherent spin frequency
$\nu$$_{pulse}$. K: van der Klis 2000; M: M\'{e}ndez et al. 1998,
M\'{e}ndez \& van der Klis 1999, 2000; B: Belloni et al. 2002, 2005;
P: Psaltis et al. 1999ab. 1: Chakrabarty 2004; 2: Linares et al.
2005; 3: Zhang et al. 2006b; 4: Wijnands et al. 2003; 5: Kaaret et
al. 2005; 6: Jonker et al. 2000; 7: O'Neill et al. 2002; 8: Jonker
et al. 2002b; 9: Homan et al. 2002; 10: Boutloukos et al. 2006; 11:
Homan 2006; 12: van Straaten et al. 2002; 13: van Straaten et al.
2000; 14: Bhattacharyya 2006; 15: van Straaten et al. 2003; 16: Di
Salvo et al. 2003; 17: Jonker et al. 2002a; 18: Markwardt et al.
1999; 19: Migliari et al. 2003; 20: Kaaret et al. 2007; 21:
Bhattacharyya et al. 2006; 22: Kaaret et al. 2002; 23: Homan and van
der Klis 2000; 24: Wijnands et al. 2001; 25: Strohmayer et al. 1997;
26: Kaaret et al. 2003.}
\end{tabular}
\end{center}
\end{table*}
%-------------------------------------------------------------------

%
\subsection{Distribution of spin frequencies in LMXBs}
In Fig.1 we plot the distribution of the 22 spin frequencies with an
average value of 440.8 Hz. For the 8 sources with both twin kHz QPOs
and spin frequencies, the ratio between the minimum upper-frequency
and the spin frequency is $\nu_{2min}/\nu_{s}>$1.3.
If the upper-frequency is interpreted as the Keplerian frequency at
the inner edge of the accretion disk (Miller et al. 1998; van der
Klis 2000; 2006), this lower limit means  that the inner edge of
disc penetrates inside the corotation radius where the Keplerian
frequency equals the spin frequency.
If this applies to the other kHz QPO sources, we can use the
relation $\nu_{s}<\nu_{2min}/1.3$ to constrain their spin
frequencies. For example, we could obtain upper limits of spin
frequencies for the 8 Z sources and 5 Atoll sources with
simultaneously detected twin kHz QPOs but with unknown spin
frequencies (see the inferred upper limits of spin frequencies for
these sources in Tab. 1).

%--------------------------------------------------------------
\begin{figure}
\centering
\includegraphics[width=\hsize]{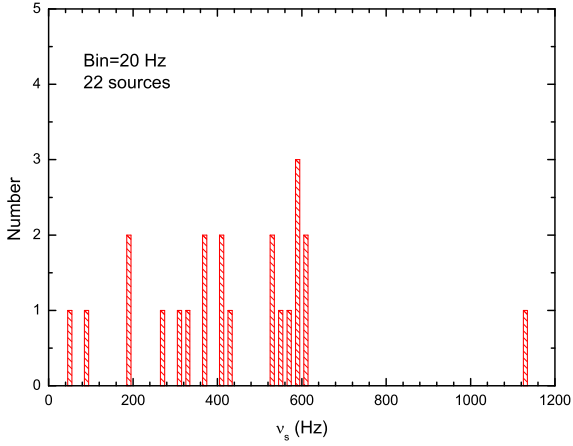}
\caption{  Distribution of the spin frequencies of the 22 neutron
stars in LMXBs.} \label{fig1}
\end{figure}
%--------------------------------------------------------------

%--------------------------------------------------------------
\begin{table}
\begin{center}
\caption{List of the sources with spin frequencies and peak
separations.}
 \label{Tab:2}
 \begin{tabular}{lccc}
\hline \hline Sources$^{*}$ & $\lan\dn\ran(\sigma)^{(1)}$ &
$\lan\dn\ran/\nu_{s}^{(2)}$ & $\nu_{2min}/\nu_{s}^{(3)}$\\
 & (Hz) & &\\
\hline {\bf Millisecond pulsar}\\
\hline
XTE J1807-294      & 215(5.9)       &  1.13 & 1.8 \\
SAX J1808.4-3658   & 195(4.0)$^{a}$ &  0.49 & 1.7 \\
\hline {\bf Atoll source}\\
\hline
4U 1608-52         &  287(7.2)     &  0.46 & 1.3 \\
4U 1636-53         &  286(9.6)     &  0.49 & 1.7 \\
4U 1702-43         &  333(8.7)     &  1.01 & 3.2 \\
4U 1728-34         &  327(5.8)     &  0.90 & 1.6 \\
KS 1731-260        &  266(8.7)     &  0.51 & 2.2 \\
4U 1915-05         &  338(12.1)    &  1.25 & 1.9 \\
\hline \hline
\end{tabular}
\begin{minipage}{\linewidth}
\begin{tabular}{p{\linewidth}}
{ $^{*}$: Data are taken from the references listed in Table 1, ;
$^{(1)}$: averaged peak separation and its standard deviation;
$^{(2)}$: ratio of averaged peak separation to spin frequency;
$^{(3)}$: ratio of the minimum upper-frequency to
 spin frequency; $^{a}$: measured  error
of the single pair of twin kHz QPOs. }
\end{tabular}
\end{minipage}
\end{center}
\end{table}
%----------------------------------------------------------------------

We also notice that, when only a single kHz QPO is detected, as in
4U 1608-52 (van Straaten et al. 2003) and 4U 1728-34 (van Straaten
et al. 2002), these QPOs do not satisfy the relation
$\nu_{s}<\nu_{2min}/1.3$. We argue that this relation only holds
when two simultaneous kHz QPOs are detected.  Our proposal would be
ruled out, if a pair of kHz QPOs were found and
$\nu_{s}>\nu_{2}/1.3$.

\subsection{Correlations between $\ns$ and $\Delta\nu$}

The sonic-point beat-frequency model (Miller et al. 1998) predicted
a constant $\dn$ equal to the stellar spin frequency, %however $\dn$
%can be less than the spin frequency about several tens of hertz
%(Lamb \& Miller 2001),
whereas the sonic-point and spin-resonance
model by Lamb \& Miller (2003) predicts that the kHz QPO peak
separation should be approximately equal to one or half the spin
frequency considering that the disk flow at the spin-resonant radius
is smooth or clumped.
%
%-------------------------------------------------------------
\begin{figure}
\centering \includegraphics[width=\hsize]{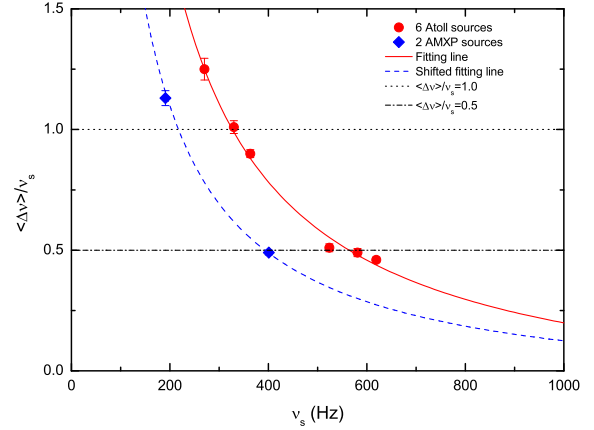} \caption{ Plot
of $\lan\dn\ran/\nu_{s}$ vs. $\nu_{s}$. The solid curve stands for
$\lan\dn\ran=-(0.19\pm0.05)\ns+(389.40\pm21.67)$Hz, and the dashed
curve is the result of shifting the solid curve down by a factor of
1.5 along the direction of the peak separation.} \label{fig2}
\end{figure}
%---------------------------------------------------------------
To confirm the above conjecture, we average  the value of peak
separations and plot $\lan\dn\ran - \ns$ diagram for the six Atoll
sources and two AMXPs in Fig. \ref{fig2}.
We notice  that for the sources (XTE J1807-294, 4U 1702-43, 4U
1728-34 and 4U 1915-05) with $\ns<400$ Hz $\lan\dn\ran/\ns \sim 1$
whereas for those (SAX J1808.4-3658, 4U 1608-52, 4U 1636-53 and 4U
1731-28) with $\ns>400$ Hz $\lan\dn\ran/\ns \sim 0.5$.
Alternatively, the relation between the averaged peak separation and
spin frequency of the six Atoll sources ($\lan\dn\ran$ and $\ns$)
can be fitted by a linear relation
$\lan\dn\ran=-(0.19\pm0.05)\ns+(389.40\pm21.67)$Hz, shown as the
solid line in Fig. \ref{fig2}. The relation indicated with a dashed
line, which crosses the points of the two AMXPs, is the same
relation as for the Atoll sources divided by 1.5 along the y
direction.

If the above anti-correlation between $\lan\dn\ran$ and $\ns$ is
real, we can use it to infer the averaged kHz QPO peak separations
of sources, such as  EXO 0748-676 ($\nu_{s}$=45 Hz), XB 1254-690
($\nu_{s}$=95 Hz) and  XTE J1739-285 ($\nu_{s}$=1122 Hz) to be
around 380 Hz, 370 Hz and 160 Hz, respectively. However, this
anti-correlation is still a conjecture since it is based on data of
only six sources. But if this result were confirmed, it means that
the spin frequency would play a role in the mechanism that produces
the kHz QPOs, but a different one from the one so far proposed.
Further measurements of kHz QPOs in the accretion powered
millisecond X-ray pulsars are required to uncover the role of the
spin of the neutron star in the mechanism that produces the kHz
QPOs.
\section{Conclusion}

Our main  conclusions are the following.

(1) We found that for the 8 sources for which twin kHz QPOs and
spins are known, the minimum upper-frequency is at least 1.3 times
larger than the spin frequency, i.e. $\nu_{2min}/\ns>1.3$. This
relation might be used to estimate the spin frequencies of sources
with twin kHz QPOs.

(2) In 6 Atoll sources, the averaged peak separation anti-correlates
with the spin frequency as
$\lan\dn\ran=-(0.19\pm0.05)\ns+(389.40\pm21.67)$Hz, although their
ratios roughly cluster around either 1 or 0.5 as reported (van der
Klis 2006). This correlation would also apply to the two AMXPs (SAX
J1808.4-3658 and XTE J1807-294) if in this cases the peak separation
is divided by a factor of 1.5 (see Fig. 2). It is noted that this
kind of shifting of about 1.5 is required to reconcile the
frequency-frequency correlation of the AMXPs and the Atoll and Z
sources (see van Straaten, van der Klis and Wijnands 2005; Linares
et al. 2005). This factor 1.5 remains unexplained, but it could
reflect a different stellar magnetic field strength or magnetic
angle between the magnetic polar axis and rotational axis between
these types of sources.

If the above correlations between the spin frequency and kHz QPOs
were confirmed in the future, it implies that the kHz QPOs are
related to the spin frequencies of neutron stars in some manner.
Thus, the spin frequency would play a role in the twin kHz QPO
production (indirectly perhaps), and any successful model of kHz
QPOs should have to take into  account these relations.
Usually,  we consider the production of  kHz QPO to be related to
the magnetosphere radius defined by the instantaneous accretion
rate. Then,  the spin frequency should be involved with the
magnetosphere radius defined by the long-term accretion rate
(matched with the different magnetic B-field when considering Atoll
and Z), which is almost stable in a short observational time.  The
instantaneous accretion rate varies around the long-term accretion
rate, which accounts for the variation of kHz QPOs, thus the
'averaged' QPO variation would be related to the spin frequency.
Therefore, in the sense of the average treatment, the kHz QPO may
have a relation to the spin frequency.

\begin{acknowledgements}
We thank T. Belloni, M. M\'{e}ndez, D. Psaltis, S. Boutloukos  and
J. Homan for providing the QPO data. Discussions with J. Petri, P.
Rebusco, J. Hor\'{a}k, V. Karas, S. Boutloukos, T.P. Li and S.N.
Zhang are highly appreciated. This research has been supported by
the innovative project of CAS of China. We are very grateful for the
critic comments from M. M\'{e}ndez.
\end{acknowledgements}

\label{lastpage}

\end{document}